\documentclass[journal]{IEEEtran}

\usepackage{xcolor,soul,framed} 

\colorlet{shadecolor}{yellow}
\usepackage[pdftex]{graphicx}
\graphicspath{{../pdf/}{../jpeg/}}
\DeclareGraphicsExtensions{.pdf,.jpeg,.png}

\usepackage[cmex10]{amsmath}
\usepackage{array}
\usepackage{mdwmath}
\usepackage{mdwtab}
\usepackage{eqparbox}
\usepackage{url}
\usepackage{hyperref}

\usepackage{ragged2e} 
\usepackage{booktabs,makecell, multirow, tabularx}
\usepackage[ruled,linesnumbered]{algorithm2e}
\usepackage[numbers]{natbib}
\SetKwComment{Comment}{// }{}

\begin{document}
\bstctlcite{IEEEexample:BSTcontrol}
    \title{Recognize then Resolve: A Hybrid Framework for Understanding Interaction and Cooperative Conflict Resolution in Mixed Traffic}
  \author{Shiyu Fang,
      Donghao Zhou,
      Yiming Cui,
      ChengKai Xu,
      Peng Hang,~\IEEEmembership{Senior Member,~IEEE,} 
      and Jian Sun

\thanks{This work was supported in part by the State Key Lab of Intelligent Transportation System under Project No. 2024-A002, the National Natural Science Foundation of China (52472451), the Shanghai Scientific Innovation Foundation (No.23DZ1203400), and the Fundamental Research Funds for the Central Universities.}
\thanks{S. Fang, D. Zhou, Y. Cui, C. Xu, P. Hang and J. Sun are with the College of Transportation, Tongji University, Shanghai 201804, China, and the State Key Lab of Intelligent Transportation System, Beijing 100088, China. (e-mail: \{2111219, zhoudonghao, 2310796, 2151162, hangpeng, sunjian\}@tongji.edu.cn).}
\thanks{Corresponding author: Peng Hang}

}

\maketitle

\begin{abstract}
A lack of understanding of interactions and the inability to effectively resolve conflicts continue to impede the progress of Connected Autonomous Vehicles (CAVs) in their interactions with Human-Driven Vehicles (HDVs). To address this challenge, we propose the Recognize then Resolve (RtR) framework. First, a Bilateral Intention Progression Graph (BIPG) is constructed based on CAV-HDV interaction data to model the evolution of interactions and identify potential HDV intentions. Three typical interaction breakdown scenarios are then categorized, and key moments are defined for triggering cooperative conflict resolution. On this basis, a constrained Monte Carlo Tree Search (MCTS) algorithm is introduced to determine the optimal passage order while accommodating HDV intentions. Experimental results demonstrate that the proposed RtR framework outperforms other cooperative approaches in terms of safety and efficiency across various penetration rates, achieving results close to consistent cooperation while significantly reducing computational resources. Our code and data are available at: \url{https://github.com/FanGShiYuu/RtR-Recognize-then-Resolve/}.
\end{abstract}


\section{Introduction}
With the advancement of autonomous driving technology, Connected and Autonomous Vehicles (CAVs) now share the roads with Human-Driven Vehicles (HDVs). However, interactions between CAVs and HDVs remain inefficient or unsafe, particularly at complex intersections. The primary challenge lies in CAVs’ insufficient understanding of interaction dynamics and their limited ability to resolve conflicts effectively. Therefore, it is imperative to thoroughly understand interaction processes and develop conflict resolution methods tailored to mixed traffic.

Modeling interaction has consistently been a prominent area of research. However, the majority of studies have focused on modeling interactive decision-making, with relatively little attention devoted to analyzing the underlying dynamics of the interaction process \cite{schwarting2019social}. Some studies have used logistics models \cite{deceunynck2013road, fang2024cooperative}, neural networks\cite{tian2022multi}, and other probabilistic models to characterize the relationship between final passage order and vehicle states, thereby enabling the prediction of interaction outcomes based on initial conditions. Other studies, from a risk quantification perspective, have identified critical moments of danger by monitoring real-time risk changes during the interaction \cite{MA2023128725, WANG2024107727}. However, the aforementioned methods often rely on threshold-based approaches to determine the final outcome, while the dynamic progression of the interaction process remains largely unexplored. Moreover, these studies primarily focus on interactions between HDVs or between HDVs and pedestrians. Consequently, their applicability to interactions between CAVs and HDVs remains limited.

Cooperative decision-making among CAVs has been shown to be a promising method for rapidly resolving conflicts, including approaches based on optimization \cite{9800916, xu2019cooperative}, game theory\cite{wang2015game,wang2021competitive,liu2024sociality}, and deep learning\cite{klimke2024towards,shi2021connected,fan2024toward}. \citet{10452809} proposed a reservation-prioritization strategy, assuming free-flowing HDV behavior, which detects conflicts between CAVs and HDVs and re-decides the passage order. \citet{le2022cooperative} estimated the Social Value Orientation (SVO) of HDVs in real time to generate balanced decisions, achieving cooperative optimal control. \citet{zhou2024reasoning} introduced a graph-based reinforcement learning method, which simplifies the solution space by incorporating physical logic relationships and predicts the potential behaviors of HDVs to generate cooperative decisions. Experimental results show a significant reduction in severe conflicts, with a 11.6\% decrease in average intersection delay. However, the aforementioned studies often treat HDVs as dynamic obstacles, which limits the full potential of cooperative decision-making. Therefore, it is essential to accurately identify the intentions of HDVs and develop conflict resolution methods suitable for mixed-traffic environments. 

Upon deeper analysis, we uncover three theoretical challenges to achieve a desirable conflict resolution: 1) \textbf{High dynamic interaction}: The interaction process is often highly volatile, requiring effective identification of moments when interactions breakdown and timely intervention in cooperative decision-making to prevent collisions or deadlocks. 2) \textbf{Complex conflict relationships}: The conflict relationships between vehicles are diverse and complex, necessitating a rapid search for the optimal cooperative passage order. 3) \textbf{Implicit HDV intentions}: Given that HDV behavior is uncontrollable, the final passage order must align with the intentions of the HDVs.
\begin{figure}[htbp]
  \begin{center}
  \centerline{\includegraphics[width=3.1in]{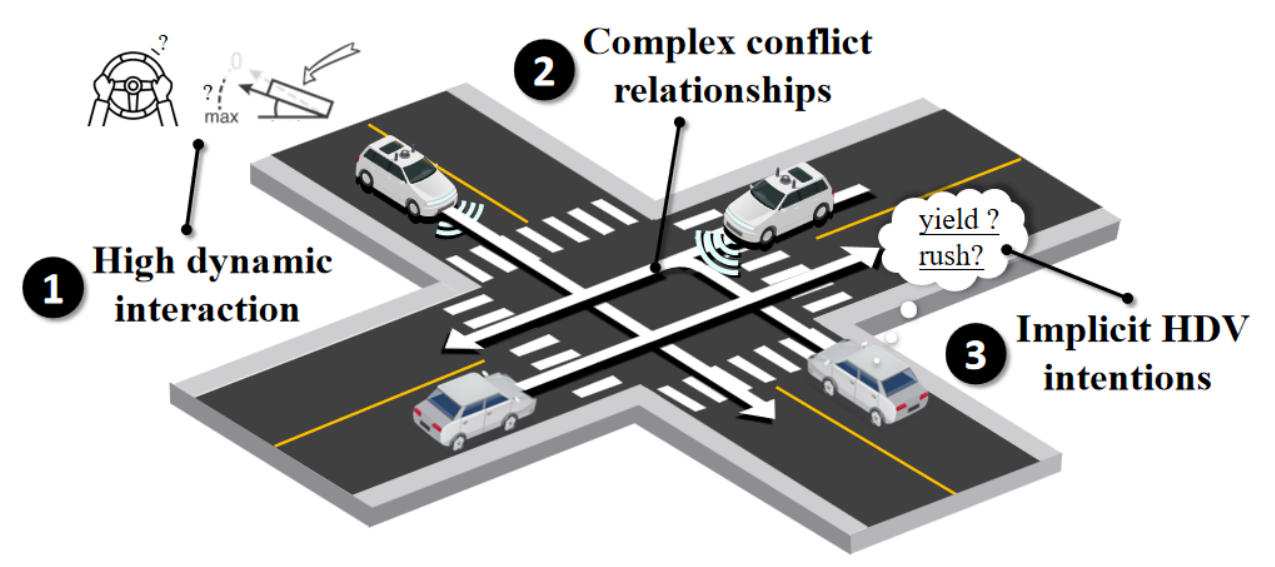}}
  \caption{Main challenges of conflict resolution in mixed traffic.}\label{challenge}
  \end{center}
  \vspace{-0.8cm}
\end{figure}

To address the aforementioned challenges, this paper proposes a Recognize then Resolve (RtR) framework for modeling the interaction process and resolving conflicts. Our contributions can be summarized as follows:

\begin{itemize}
    \item Proposal of a Bilateral Intention Progression Graph (BIPG) to model the development of interactions based on CAV-HDV interaction data and to infer HDV intentions.
    \item Definition of three typical types of interaction breakdown scenarios to identify critical moments for triggering cooperative conflict resolution.
    \item Introduction of a constrained Monte Carlo Tree Search (MCTS) algorithm to optimize the passage order while maintaining alignment with HDV intentions.
    \item Validation of the proposed RtR framework under varying penetration rates, demonstrating superior safety and efficiency compared to other methods, while significantly reducing computational resources and achieving performance close to consistent cooperation.
\end{itemize}

\section{Problem Formulation}
\subsection{Assumptions}
To simplify the problem and focus on the core aspects of cooperative conflict resolution at mixed un-signalized intersections, the following assumptions are made:
\begin{itemize}
    \item Lane-changing maneuvers upstream of the intersection are excluded.
    \item Information (positions, speeds) about vehicles within a certain range of the intersection can be obtained.
    \item The intention of HDV is simplified as either rush or yield.
\end{itemize}

\subsection{Formulation of the Cooperative Conflict Resolution}
The primary aim of the proposed method is to search for the optimal passing order \(O^*\) for vehicles at an intersection that aligns with the intentions of HDVs. This problem can be modeled as a constraint-aware sequential decision-making process that follows:
\begin{equation}
\begin{split} \label{formulation}
& O^* =\arg\max_O\sum_{i=1}^NR(S_i,o_i)\\
s.t. \;\;&  o_h = I_h = \left\{
\begin{array}{ll}
0, \text{rush} \\
1, \text{yield}
\end{array}
\right.\; ,\forall\; h \in H
\end{split}
\end{equation}
where \(N\) denotes the set of all vehicles involved in the interaction. \(H \subseteq N\) represents the subset of HDVs. For any \(h \in H\), \(I_h\) refers to the intended passing order of HDV \(h\). The reward function \(R(S_i, o_i)\) evaluates the quality of the passing order \(o_i\) for vehicle \(i\) in state \(S_i\). The objective is to maximize the cumulative reward across all vehicles, while strictly respecting the intentions of HDVs.

\section{Methodology}
\begin{figure*}[htbp]
  \begin{center}
  \centerline{\includegraphics[width=7in]{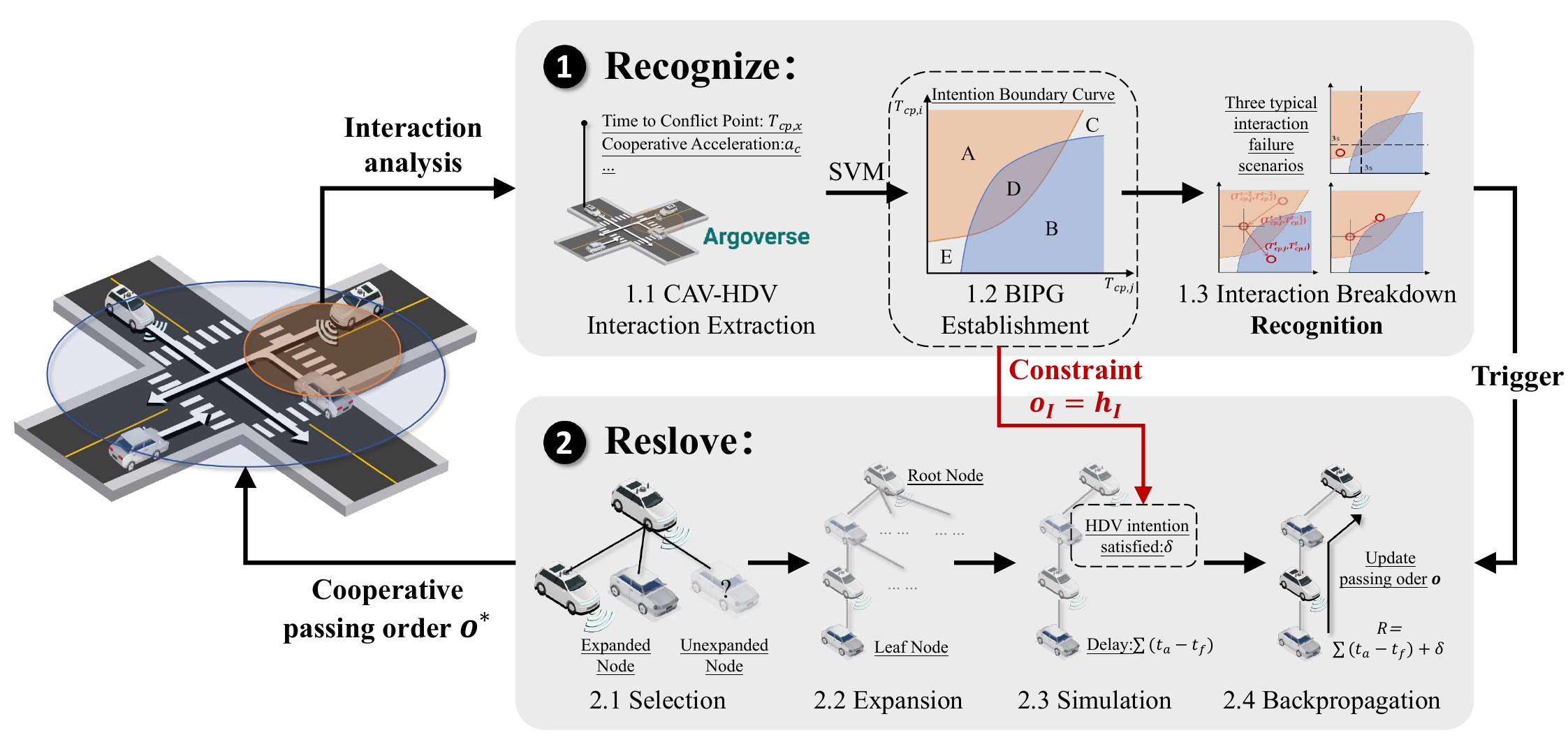}}
  \caption{Overview of the proposed RtR framework for interaction breakdown recognition and conflict resolution.}\label{framework}
  \end{center}
  \vspace{-0.8cm}
\end{figure*}

To model the interaction process, identify the latent intentions of HDVs, and search for the optimal cooperative passing order, this paper proposes the RtR framework, as illustrated in Fig.~\ref{framework}. Initially, a BIPG is established to identify interaction breakdown and the intentions of HDVs. Additionally, the MCTS algorithm is utilized to determine the optimal passing order while ensuring the HDV intention constraints are met.

\subsection{\textbf{Recognize}: Identify Interaction Breakdown with Bilateral Intention Progression Graph}

To identify the point at which interactions begin to breakdown, this section introduces a BIPG to describe the development of the interaction process. First, a total of 901 interaction samples between CAVs and HDVs were extracted from the Argoverse dataset. Additionally, to quantify the dynamic interplay between participants during the interaction, cooperative acceleration \(a_c\) is introduced, as defined in Eq.~\ref{cooperative acceleration}. This metric evaluates the minimal potential impact of a vehicle’s actions on its counterpart, considering the interaction from the perspective of both participants.
\begin{equation}
\begin{split} \label{cooperative acceleration}
& a_{c,i}=\frac{2(d_{i}-v_{i}T_{cp,j})}{T_{cp,j}^2} \\
& T_{cp} = \frac{d}{v}
\end{split}
\end{equation}
where \(d_{i}\) is the distance between interaction opponent vehicle \(i\) and the collision point, and \(v_{i}\) is its current speed, \(T_{cp,j}\) is the Time-to-Collision Point (TTCP) for ego vehicle \(j\). 

To explore the relationship between vehicle states and the final passing order, we formulate the passing decision as a classification problem using the Support Vector Machine (SVM), with the possible outcomes being rush or yield as described in Eq.~\ref{formulation}. Through experimentation, we discovered that incorporating TTCP and cooperative acceleration as features delivers optimal performance. Based on this, the SVM formulation can be derived:
\begin{equation}
\begin{split} \label{svm}
&\boldsymbol{\omega^T}\boldsymbol{X}+b=0 \\
&\omega_1T_{cp,i}+\omega_2T_{cp,j}+\omega_3a_c,j+b=0
\end{split}
\end{equation}
where \( \boldsymbol{\omega} \) is the vector of weights that determines the orientation of the hyperplane in the feature space, \( \boldsymbol{X} \) is the input feature vector, \( b \) is the bias term.

Substituting Eq.~\ref{cooperative acceleration} into Eq.~\ref{svm} and rearranging the terms, the following expression is obtained:
\begin{equation}
\label{svm2}
T_{cp,i}= -\frac{\omega_{1}}{\omega_{2}} T_{cp,j} -\frac{\omega_{3}}{\omega_{2}} \frac{d_s-v_sT_{cp,j}}{T_{cp,j}^2} - \frac{b_l}{\omega_{2}}
\end{equation}

Based on Eq.~\ref{svm2}, the intention boundary curve of both interacting vehicles can be plotted as shown in Fig.~\ref{framework} (1.2). The potential intention of the interacting vehicles can be determined based on the current interaction state and the region it falls into.

The intention boundary curve delineates the interaction process into five distinct regions: 1) \textbf{Region A}: Vehicle \( j \) prefers to proceed first. 2) \textbf{Region B}: Vehicle \( i \) prefers to proceed first. 3) \textbf{Region C}: Both vehicles prefer to proceed first, but the distance to the conflict point is sufficiently large, allowing ample time for adjustments. 4) \textbf{Region D}: Both vehicles prefer to yield, resulting in a deadlock. 5) \textbf{Region E}: Both vehicles prefer to proceed first.

\begin{figure}[htbp]
  \begin{center}
  \centerline{\includegraphics[width=3.8in]{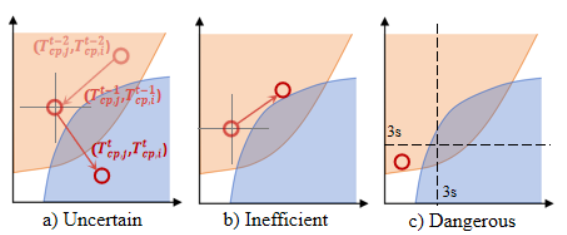}}
  \caption{Illustrations of three typical interaction breakdown scenarios.}\label{intention}
  \end{center}
  \vspace{-0.8cm}
\end{figure}

Additionally, three typical interaction breakdown scenarios are identified: uncertain, inefficient, and dangerous interactions, as shown in Fig.~\ref{intention}. Uncertain interaction occurs when the convergence direction changes during the process, causing fluctuations in the expected passing order. Inefficient interaction is marked by simultaneous increases in the TTCP values of both vehicles, potentially leading to a deadlock. Dangerous interaction happens when the TTCP values of both vehicles fall below 3 seconds, indicating a high collision risk. When any of the three aforementioned phenomena occur, it can be inferred that the interaction is progressively failing, necessitating the intervention of cooperative decision-making to enhance safety and efficiency.

\subsection{\textbf{Resolve}: Searching Optimal Passing Order with MCTS}

Once the moment requiring cooperative driving is identified, MCTS is introduced to determine the optimal passing order. During this process, the intentions of HDVs are incorporated as constraints to prevent conflicts or deadlocks caused by mismatches between the final passing order and driver intentions. In general, MCTS consists of four main steps: Selection, Expansion, Simulation, and Backpropagation.

\subsubsection{Selection}
In the selection phase, starting from the root node, the algorithm selects child nodes sequentially based on a balance between exploration and exploitation, as described by Eq.~\ref{ucb}, until an unexpanded node is reached. Each node represents the state of a vehicle, and the path from the root node to the leaf node forms the passing order of vehicles. 
\begin{equation}
\label{ucb}
\arg \max _{k} \bar{Q}_{k}+C \sqrt{\frac{2 \ln n_{p}}{n_{k}}}
\end{equation}
where \( \bar{R}_{k} \) is the average reward of node \(k\), representing the utility of this passing order, \( C \) is a hyper-parameter used to balance exploration and exploitation, \( n_{p} \) and \( n_{k} \) are the number of times the parent node \( p \) and current node \( k \) was visited. The average reward can be caculated as follows:
\begin{equation}
\label{reward}
\bar{R}_{k}=\frac{1}{n_{k}} \sum_{n=1}^{n_{k}} R_{k,n} = \frac{1}{n_{k}} \sum_{n=1}^{n_{k}}\sum_{d=1}^{D} (t_{a,d} - t_{f,d})
\end{equation}
where \(R_{k,n}\) is the reward when node \(k\) is visited for the d-th time, \(t_{f,i} \) and \(t_{a,i} \) are the free flow time and actual travel time of vehicle \(i\), \(D\) denotes the set of vehicles that have been assigned a passing order in the current exploration.

\subsubsection{Expansion}
If the generated passing order does not include all vehicles, the process continues to expand according to Eq.~\ref{ucb} to determine the next vehicle's passage order. Once all vehicles have been assigned a passage order, the simulation phase begins.

\subsubsection{Simulation}
The Simulation phase evaluates the cumulative reward \(R(o) \) based on the passing order \(o \) formed from the root node to the leaf node as follows:
\begin{equation}
\label{sum reward}
\begin{split}
&R(o)= \sum_{k=1}^{N}\bar{R}_{k}+\lambda\sum_{h\in H}\delta(o_h,I_h)\\
&\delta(o_h,I_h)=\left\{
\begin{array}{ll}
0,\; o_h = I_h \\
1,\; else
\end{array}
\right.
\end{split}
\end{equation}
where \(\lambda\) is a penalty weight to avoid the actual passing order conflicting with the HDV's intentions, \(\delta \) is the Kronecker delta function.

\subsubsection{Backpropagation}
Finally, the cumulative reward \(R(o) \) from the simulation is propagated back through the nodes along the passing order to the root. The statistics of each visited node are updated as follows:
\begin{equation}
\label{update n}
n_{k} = n_{k}+1
\end{equation}

By iteratively executing these steps, MCTS progressively refines the passing order and generates the optimal passing order that aligns with HDVs' intentions.

\section{Experiments and Analysis}

\subsection{Experiment Settings}
\textbf{Reproduction of HDV.} Reproducing real-world heterogeneous HDVs is crucial for evaluating the RtR framework. Based on our previous research \cite{fang2024cooperative}, drivers were categorized as aggressive, conservative, or normal. Maximum entropy inverse reinforcement learning was then applied to calibrate their decision preferences for safety, efficiency, and comfort. Finally, A non-cooperative Bayesian model was used to simulate the decisions of heterogeneous HDVs.

\textbf{Implementation details.} Each experiment was conducted at penetration rates of 30\%, 50\%, 70\%, and 100\%, with initial positions, velocities, and driving styles of HDVs randomly generated. Each condition was simulated 100 times. When a CAV operated in single-vehicle decision mode, IDM was used as the underlying decision-making model. During cooperation, vehicles with higher priority were treated as the lead vehicle in IDM to generate acceleration. More detailed explanations, along with the video and raw data, are available here \footnote{\url{https://fangshiyuu.github.io/RtR-Recognize-then-Resolve/}}.

\subsection{Validation of BIPG through CAV Takeover Case}
\begin{figure}[tbp]
  \begin{center}
  \centerline{\includegraphics[width=3.2in]{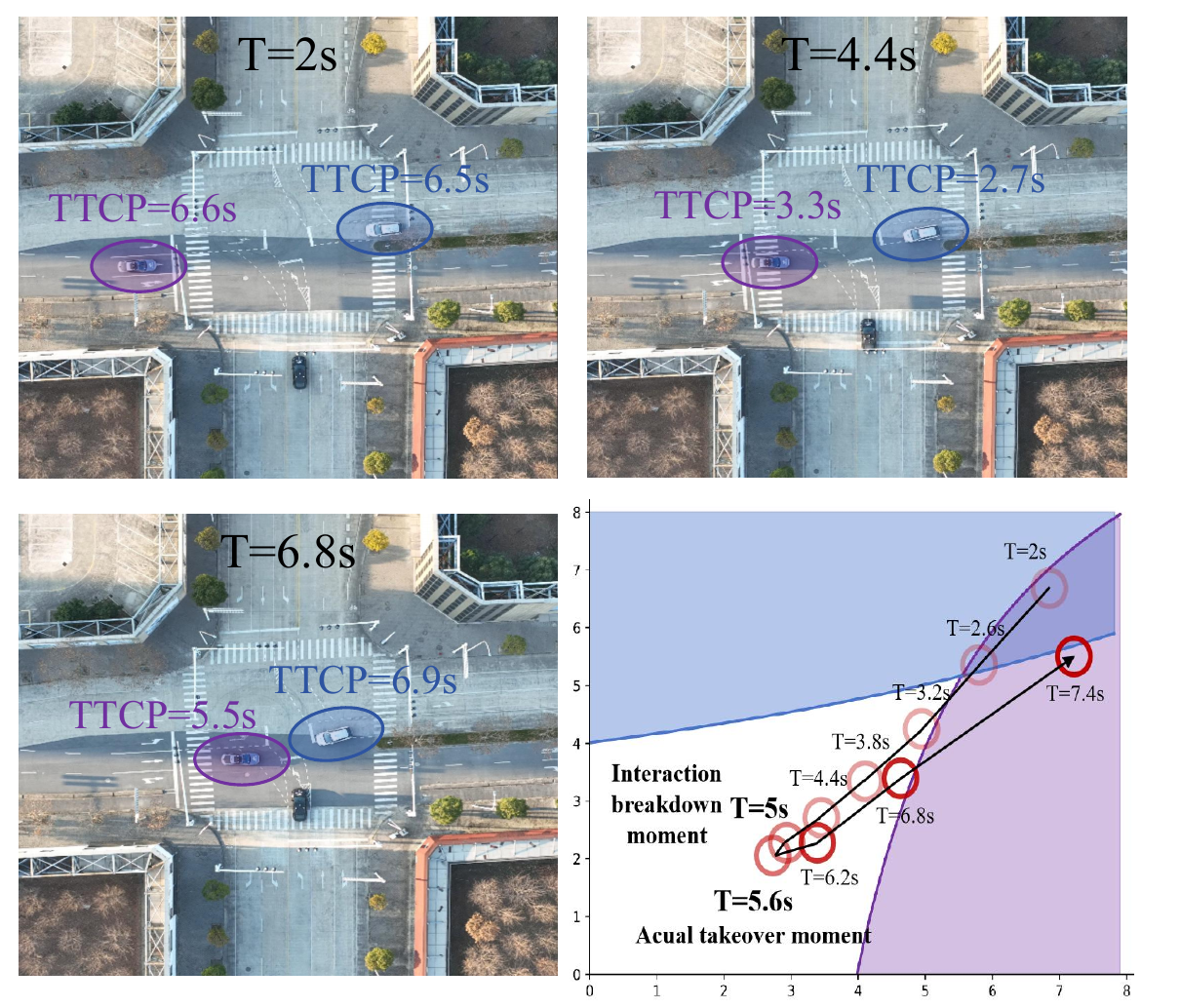}}
  \caption{A CAV takeover case while interacting with HDV at un-signalized intersection.}\label{deadlock_case}
  \end{center}
  \vspace{-0.8cm}
\end{figure}

First, to validate the effectiveness of the proposed interaction breakdown recognition based on BIPG, we analyzed a case where a CAV interacted with an HDV at an open un-signalized intersection. Fig.~\ref{deadlock_case} illustrates key moments during the interaction, showing the vehicles' real-time positions and the state changes in BIPG at different timestamps. At \(T=5\)s, both the CAV and HDV had TTCP values below 3s, indicating a dangerous interaction requiring intervention to ensure safety. Approximately 0.6s later, the CAV driver manually took over and applied the brakes. This demonstrates that the proposed BIPG-based method aligns closely with the driver’s actual intention during dynamic interactions and can even recognize potential risks earlier, enabling timely intervention.

\subsection{Ablation Study of RtR Framework}
\begin{figure}[htbp]
  \begin{center}
  \centerline{\includegraphics[width=3.2in]{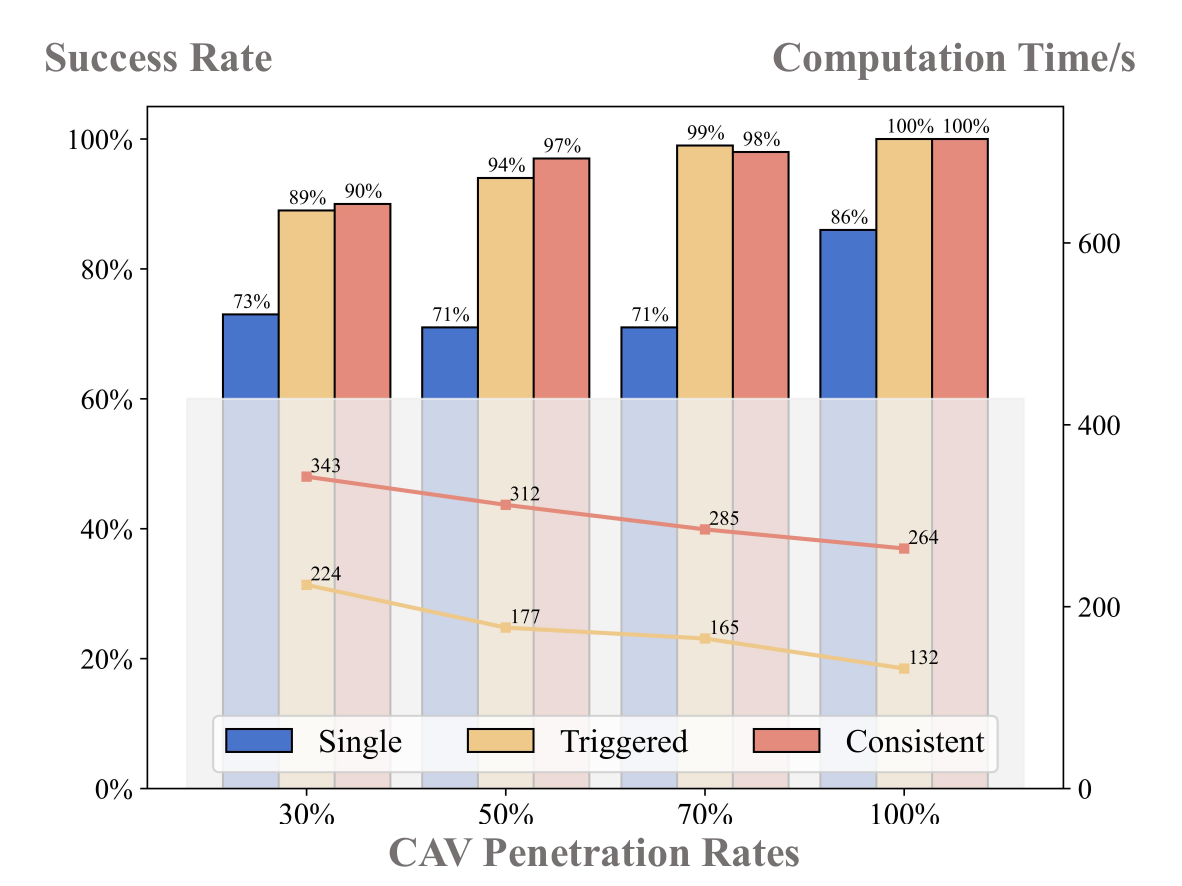}}
  \caption{Variations in success rate and computation time under different penetration rates.}\label{ablation}
  \end{center}
  \vspace{-0.8cm}
\end{figure}

Moreover, to evaluate the effectiveness of the RtR framework, we compared the success rates and computation times of single-vehicle decision-making, consistent cooperation, and triggered cooperation under different penetration rates, as shown in Fig.~\ref{ablation}. In each simulation, success is defined as all vehicles leaving the intersection without collisions within 30s. 

Specifically, single-vehicle decision-making refers to all CAVs using the IDM model. Consistent cooperation involves solving the optimal cooperative passing order using MCTS at every simulation frame. Triggered cooperation, which represents the proposed RtR framework, involves applying cooperative decision-making only during interaction breakdown while relying on single-vehicle decision-making at all other times.
\begin{figure*}[t]
  \begin{center}
  \centerline{\includegraphics[width=7in]{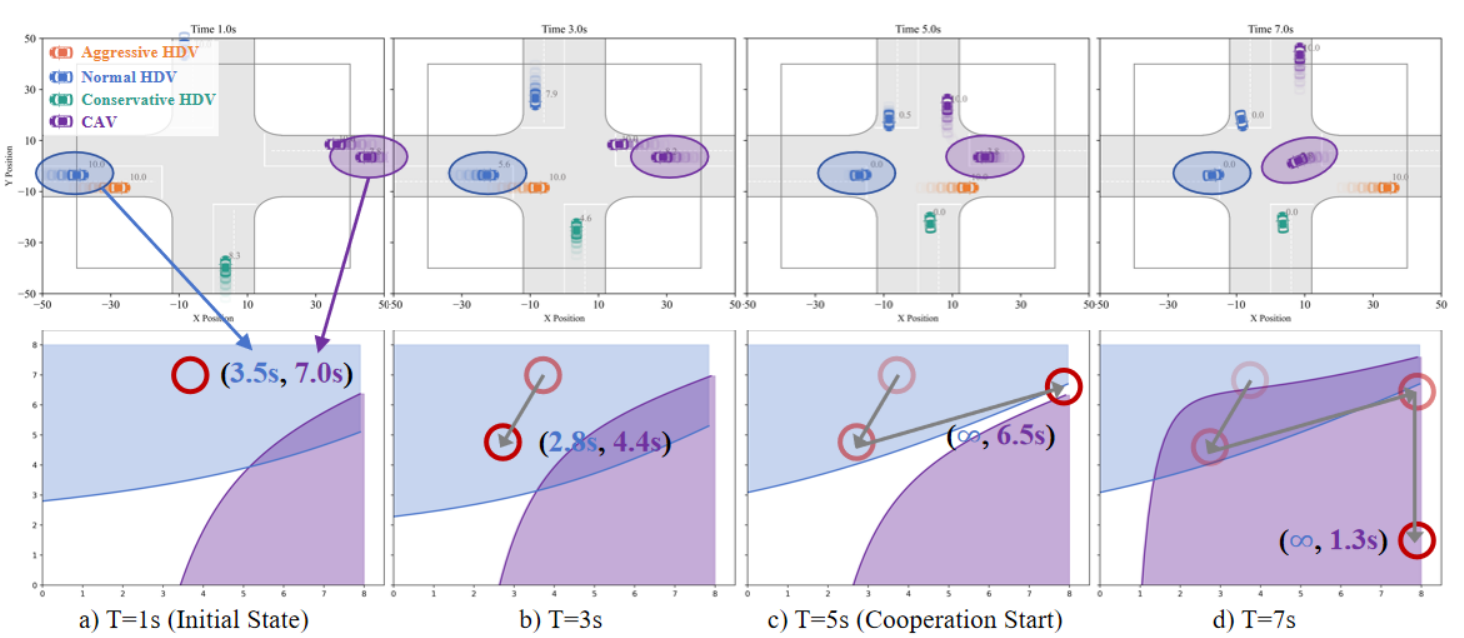}}
  \caption{A real-time vehicle cooperation case driven by the proposed RtR framework.}\label{case}
  \end{center}
  \vspace{-0.8cm}
\end{figure*}

As shown in Fig.~\ref{ablation}, the success rates of both triggered and consistent cooperation steadily improves with increasing CAV penetration rates, significantly outperforming single-vehicle decision-making. Triggered cooperation achieves an average success rate reduction of less than 1\%, with a maximum decrease of only 3\%.

Regarding computation time, both cooperation types decrease as the CAV penetration rate increases, with triggered cooperation consistently requiring less time. At 100\% penetration, its computation time is half that of consistent cooperation. Therefore, the RtR framework demonstrates performance nearly identical to consistent cooperation while significantly reducing computational resources, highlighting the importance of recognizing interaction breakdown moments.

Additionally, Fig.~\ref{case} illustrates a real-time vehicle cooperation process within the RtR framework. Initially, the HDV aims to pass the intersection first, maintaining this intention up to $T=3s$. However, at $T=5s$, the HDV unexpectedly decides to yield to the CAV, leading to inefficient interaction and therefore triggering cooperation. At this point, the HDV’s yielding intent is incorporated as a constraint into the MCTS. After computation, the optimal passing sequence prioritizes the CAV. Subsequently, at $T=7s$, the CAV accelerates and clears the intersection.

\subsection{Comparison with Other Method}
Finally, to evaluate the effectiveness of the proposed RtR framework, we compare its performance in terms of overall success rate, safety, and efficiency against First-Come-First-Serve (FCFS) \citep{dresner2008multiagent}, improved Depth-First Search Tree (iDFST) \citep{xu2018distributed, chen2022conflict}, and Cooperative Game (CGame) \citep{fang2024cooperative} approaches under varying penetration rates.

\subsubsection{Overall Performance}
Table.~\ref{tab:performance} highlights the success rate variations across different methods under varying penetration rates. The proposed RtR framework consistently outperforms other methods at all penetration levels, achieving a success rate of nearly 89\% even with only 30\% CAV penetration. 

In contrast, FCFS and iDFST, as reservation-based approaches, show limited success rates of around 70\% in mixed traffic with HDVs present. However, their success rates rapidly reach 100\% in fully CAV environments, underscoring the limitations of reservation-based methods in mixed traffic scenarios. Lastly, while CGame performs better overall than reservation-based approaches, its assumption that all agents are cooperative may neglect the heterogeneity of real-world drivers. As a result, its success rates remain consistently lower than those of the proposed RtR framework across all penetration levels.
\begin{table}[h]
    \caption{Success rate under various methods}
    \centering
    \centering
    \begin{tabular}{c| c c c c}
    \hline
     \multirow{2}{*}{Penetration rates} & \multicolumn{4}{c}{Method} \\
     \cline{2-5}
       &  RtR & FCFS & iDFST & CGame  \\
     \hline
     30\% &  \textbf{89\%} & 70\% & 72\% & 76\% \\
     50\%  & \textbf{94\%} & 64\% & 74\% & 87\% \\
     70\%  & \textbf{99\%} & 71\% & 74\% & 87\%  \\
     100\%  & \textbf{100\%} & 100\% & 100\% & 94\% \\
    \hline
\end{tabular}
    \label{tab:performance}
\end{table}

\subsubsection{Safety Evaluation}
Furthermore, we conducted a deeper comparison of safety performance using Post-Encroachment Time (PET) analysis, where collision cases were excluded to better assess the risk levels of non-colliding interactions. Fig.~\ref{case} depicts the PET distribution for different methods at a 70\% penetration rate. Although the differences in average PET are not substantial, the proportion of dangerous interactions with PET less than 3 seconds is notably smaller for the proposed RtR framework compared to the other methods. Additionally, PET distributions for 30\%, 50\%, and 100\% penetration rates can also be visualized, though due to space limitations, these are not presented here; they can be found on the project homepage. Overall, the RtR framework consistently demonstrates a low occurrence of dangerous interactions across all penetration levels compared to the other cooperative methods.
\begin{figure}[htbp]
  \begin{center}
  \centerline{\includegraphics[width=3.2in]{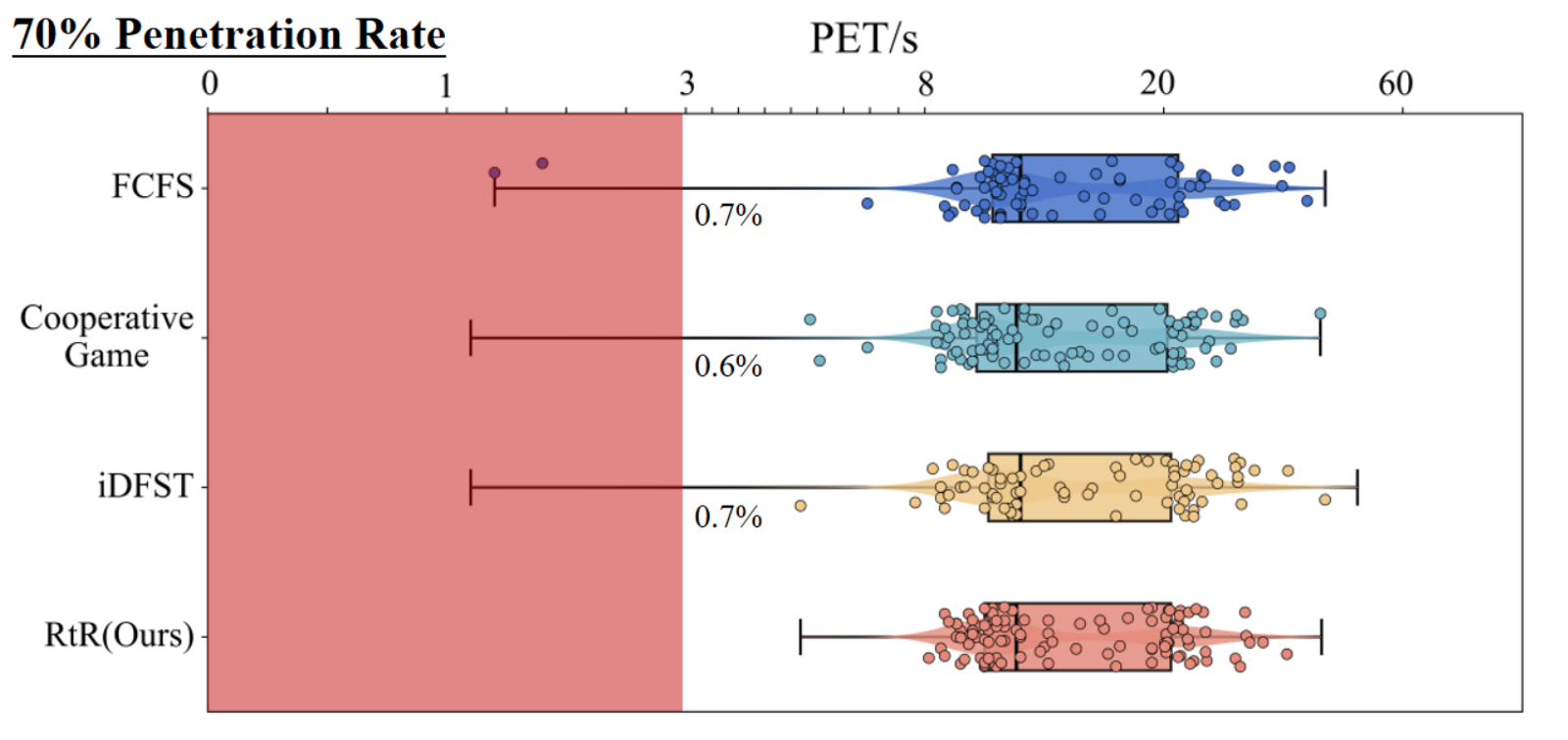}}
  \caption{PET distribution of various cooperative method under 70\% CAV penetration rate.}\label{pet}
  \end{center}
  \vspace{-0.8cm}
\end{figure}

\subsubsection{Efficiency Evaluation}
\begin{figure}[htbp]
  \begin{center}
  \centerline{\includegraphics[width=2.8in]{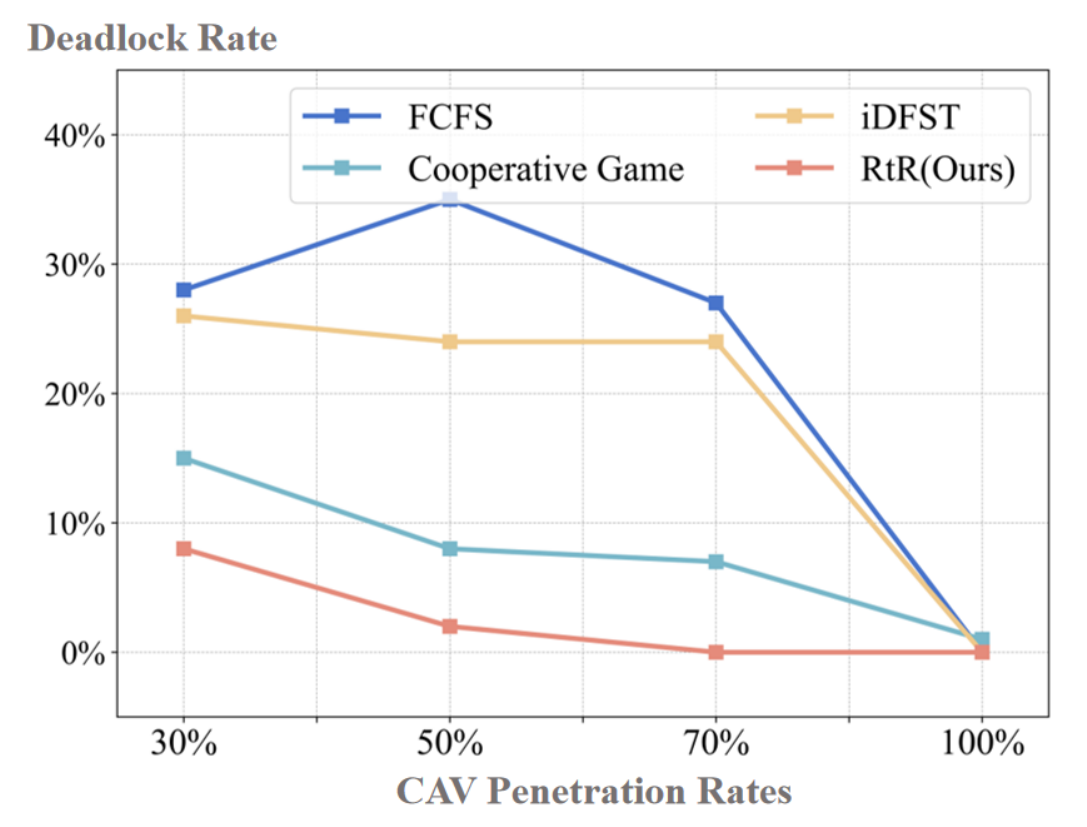}}
  \caption{Deadlock rate under different penetration rates.}\label{deadlock_rate}
  \end{center}
  \vspace{-0.8cm}
\end{figure}

Finally, we compared the efficiency of different methods using the deadlock rate, as shown in Fig.~\ref{deadlock_rate}. The proposed RtR framework demonstrated the best performance in terms of efficiency, followed by CGame. In contrast, reservation-based methods, such as FCFS and iDFST, showed the poorest performance. This inefficiency can be attributed to their reliance on predefined rules, which are not adaptive to the uncertain mixed traffic.

Overall, the RtR framework shows clear advantages in both efficiency and safety across all penetration rates and consistently maintains strong performance even in environments with heterogeneous HDVs.

\section{Conclusion}
Currently, interactions between CAVs and HDVs remain inefficient and unsafe, primarily because CAVs struggle to accurately interpret the dynamics of interaction processes. Moreover, resolving conflicts effectively often requires cooperation among multiple vehicles rather than relying solely on the efforts of a single vehicle. To address these challenges, this study proposes a Recognize then Resolve (RtR) framework. The framework begins by constructing a BIPG to identify the HDV's intention and the moments of interaction breakdown. It then uses a constrained MCTS to determine optimal cooperative passing order to resolve conflict. 

To validate the proposed method, we analyzed a real-world CAV takeover case and demonstrated that BIPG enables earlier detection of interaction breakdowns. Additionally, ablation experiments showed that the RtR framework achieves similar performance to consistent cooperation while reducing computational resources by approximately 50\%. Finally, comparative analyses confirmed that the RtR framework outperforms existing methods in terms of both safety and efficiency.

%
\IEEEpeerreviewmaketitle

\ifCLASSOPTIONcaptionsoff
  \newpage
\fi

\footnotesize
\bibliographystyle{IEEEtranN}
\bibliography{IEEEabrv,Bibliography}

\vfill

\end{document}